\documentclass[twocolumn,showpacs,prb]{revtex4}
\usepackage{graphicx}

\newcommand{\BE}{\begin{equation}}
\newcommand{\BEA}{\begin{eqnarray}}
\newcommand{\EE}{\end{equation}}
\newcommand{\EEA}{\end{eqnarray}}
\newcommand{\NN}{\nonumber}
\renewcommand{\r}{\right}
\renewcommand{\l}{\left}
\newcommand{\R}{\rangle}
\renewcommand{\L}{\langle}
\renewcommand{\a}{\alpha}
\renewcommand{\b}{\beta}
\newcommand{\g}{\gamma}
\newcommand{\s}{\sigma}
\newcommand{\om}{\omega}
\newcommand{\eps}{\varepsilon}
\renewcommand{\d}{{\rm d}}
\newcommand{\e}{{\rm e}}
\newcommand{\sgn}{{\rm sign}}
\newcommand{\bra}[1]{\langle #1|}
\newcommand{\ket}[1]{|#1\rangle}
\newcommand{\Gc}{{\cal G}}

\newcommand{\Om}{\Omega}
\newcommand{\dg}{\dagger}
\newcommand{\Vh}{{\hat V}}
\newcommand{\Gh}{{\hat G}}
\newcommand{\up}{\uparrow}
\newcommand{\dn}{\downarrow}
\newcommand{\sumind}[2]{{\scriptstyle #1\atop\scriptstyle #2}}
\newcommand{\dk}[1]{{\d #1\over(2\pi)^d}}
\newcommand{\dom}{{\d\om\over2\pi}}
\newcommand{\norb}{n_{\rm orb}}
\newcommand{\rv}{{\bf r}}
\newcommand{\Rv}{{\bf R}}
\newcommand{\Qv}{{\bf Q}}
\newcommand{\Kv}{{\bf K}}
\newcommand{\kv}{{\bf k}}
\newcommand{\qv}{{\bf q}}
\newcommand{\DO}{{\cal D}}
\newcommand{\Ecin}{\eps_{\rm kin}}
\newcommand{\Epot}{\eps_{\rm pot}}
\newcommand{\Imag}{{\,\rm Im\,}}
\newcommand{\At}{{\tilde A}}
\begin{document}

\title{Cluster Perturbation Theory for Hubbard models}
\author{David S\'en\'echal}
\email{david.senechal@courrier.usherb.ca}
\author{Danny Perez}
\author{Dany Plouffe}
\affiliation{D\'epartement de physique, Universit\'e de Sherbrooke, Qu\'ebec, Canada J1K 2R1}

\begin{abstract}
Cluster perturbation theory is a technique for calculating the spectral weight of Hubbard models of strongly correlated electrons, which combines exact diagonalizations on small clusters with strong-coupling perturbation theory at leading order. It is exact in both the strong- and weak-coupling limits and provides a good approximation to the spectral function at any wavevector. Following the paper by S\'en\'echal et al. (Phys. Rev. Lett. {\bf 84}, 522 (2000)), we provide a more complete description and derivation of the method. We illustrate some of its capabilities, in particular regarding the effect of doping, the calculation of ground state energy and double occupancy, the disappearance of the Fermi surface in the $t-t'$ Hubbard model, and so on. The method is applicable to any model with on-site repulsion only.
\end{abstract}

\maketitle


\section{Introduction}

It is generally agreed that correlation effects play a central role in the Physics of many classes of materials, including high-temperature superconductors, manganites and organic superconductors. Unfortunately, making definite predictions from theoretical models of strongly correlated electrons is notoriously difficult, because these systems are not well served by the usual approximation schemes of many-body quantum mechanics. 
But ARPES experiments, which have improved steadily in resolution over the past decade\cite{Shen95, Damascelli01}, clearly show the need to go beyond simple band-structure interpretations of the results.

It is generally taken for granted that ARPES experiments measure $f(\om)A(\kv,\om)$, where $A(\kv,\om)$ is the spectral function, the probability for an electron of wavevector $\kv$ to have an energy $\hbar\om$, and $f(\om)$ is the Fermi-Dirac function. At zero temperature, this boils down to the negative-frequency part of $A(\kv,\om)$. In fact, things are more complicated: $A(\kv,\om)$ is the density of final states in Fermi's golden rule, and the matrix element may have a non-negligible frequency and momentum dependence, even though it is commonly neglected. Nevertheless, even $A(\kv,\om)$ can be a challenging object to calculate within a correlated electron model.

\begin{figure}
\centerline {\includegraphics[width=6cm]{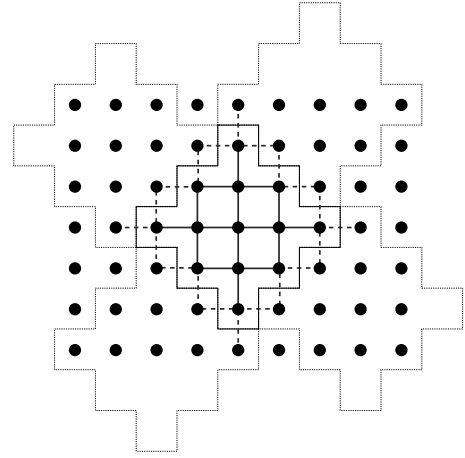}}
\caption{Tiling of the square lattice with 13-site clusters. The NN hoppings within a cluster are represented by full lines, and inter-cluster hoppings by dashed lines. Of course, much simpler tilings (e.g. square) are possible.}
\label{A13-FIG}
\end{figure}

In this paper, we will explain in detail a technique introduced in Ref.~\onlinecite{Senechal00}, named Cluster Perturbation Theory (CPT), that can be applied to Hubbard models (i.e. models of correlated electrons with on-site interaction) in order to calculate the spectral function $A(\kv,\om)$.
The basic idea behind cluster perturbation theory is to divide the lattice into a superlattice of identical clusters (Fig.~\ref{A13-FIG}). The Hubbard model on each cluster is solved exactly (the electron Green function is calculated numerically), whereas the hopping between sites belonging to different clusters is treated perturbatively. This paper is methodological in character; even though results relevant to actual materials are presented (e.g. Fig.~\ref{pseudo-FIG}), the emphasis is placed on the method itself.

Let $\g$ denote the original Bravais lattice of the model, and $\Gamma$ the superlattice of clusters, such that $\Gamma\subset\g$; the cluster is formally the quotient $\g/\Gamma$. We will use the notation $\rv$ and $\Rv$ for the lattice sites of $\g$ and $\Gamma$, respectively, and $L$ will denote the number of lattice sites within a cluster. An example of clustering of the two-dimensional square lattice with $L=13$ is shown on Fig.~\ref{A13-FIG}. Tiling the lattice with rectangular clusters is obviously simpler, but being able to compare the results of different tilings is important, in order to separate real physical effects from artefacts caused by a particular cluster shape. Each lattice site $\rv$ of a cluster can also host a number $\norb>1$ of orbitals (as in a many-band model), although the case $\norb=1$ will be most frequent.

The class of Hamiltonians that can be treated by this approach has the form $H=H_0+V$, where
\BE\label{classH}
H_0 = \sum_\Rv H^0_\Rv \quad,\quad
V = \sum_\sumind{\Rv,\Rv'}{a,b} V^{\Rv,\Rv'}_{a,b} c_{\Rv a}^\dg c_{\Rv' b}
\EE
Here $a,b$ label different electron orbitals (or sites) within a cluster, and range from 1 to $\norb L$. $H^0_\Rv$ is a Hubbard Hamiltonian defined on a single cluster and $V$ is a hopping term between clusters, $V^{\Rv,\Rv'}_{a,b}$ being the hopping amplitude between orbital $a$ of cluster $\Rv$ and orbital $b$ of cluster $\Rv'$. Hopping can be of any range, and multi-band models are treated on the same footing as the one-band model: The indices $a,b$ may refer to different lattice sites on a cluster, or to different bands (orbitals) within a cluster, or both.  In order to stay as general as possible, we will call them ``orbital indices''. Spin indices are implicit in hopping terms and suppressed where appropriate in order to lighten the notation. 

In the simplest case, treated explicitly in Refs~\onlinecite{Pairault98,Pairault00}, the cluster consists of a single site ($\g=\Gamma$ and $\norb=1$) and the perturbation $V$ is a nearest-neighbor hopping~:
\BE
H^0_\rv = Un_{\rv,\up} n_{\rv,\dn} \quad,\quad
V^{\rv,\rv'}  = -t \quad\text{($\rv,\rv'$ N.N.)},
\EE
In that case, the perturbative treatment of $V$ coincides with the so-called strong-coupling perturbation theory (SCPT) of the Hubbard model and has been extensively studied in Ref.~\onlinecite{Pairault00}. Additional references on the strong-coupling perturbation theory, in particular earlier attempts at its systematization, can be found therein. We review the main results in the following subsection. In practice,  SCPT provides approximate expressions for the various electron Green functions in terms of the hopping integral and of various atomic Green functions (i.e., Green functions of the one-site Hubbard model). The general results of SCPT can be immediately applied to the class (\ref{classH}) of Hamiltonians, with the difference that atomic Green functions are replaced with exact Green functions of a Hubbard model defined on a small cluster. 

In Sect.~\ref{RemarksS}, we make a series of important remarks on the method.
In Sect.~\ref{clusterS} we explain how the Green function on the cluster is calculated numerically. We comment on multi-band models in Sect.~\ref{multibandS}. We conclude in Sect.~\ref{energyS} with a calculation of integrated quantities, like the energy density and double occupancy.

\section{Derivation of the method}

\subsection{The strong-coupling expansion}
\label{SCPT-SS}

In this section we review the main results of Ref.~\onlinecite{Pairault98,Pairault00} on strong-coupling perturbation theory, in a notation adapted to the present situation. It is assumed that we are dealing with a Hamiltonian of the form (\ref{classH}). Our goal is to place the approximation made in CPT in the context of a systematic strong-coupling perturbation theory, i.e. to explain in what way it constitutes a leading order approximation.

In the path-integral formalism, the partition function is expressed as a functional integral over Grassmann variables $\g_{\Rv a}$ that correspond to the creation operator $c_{\Rv a}$~:
\begin{widetext}
\BE
Z=\int [d\g^\star d\g] \exp-\int_0^\b d\tau \l\{ 
\sum_{\Rv,a}\g_{\Rv a}^\star(\tau)\l( \partial_\tau -\mu \r)
\g_{\Rv a}(\tau) + \sum_\Rv H^0_\Rv (\g_{\Rv a}^\star,\g_{\Rv a}) + \sum_\sumind{\Rv,\Rv'}{a,b} V^{\Rv,\Rv'}_{a,b} \g_{\Rv a}^\star \g_{\Rv' b}
\r\}
\EE
In order to lighten the notation, we use greek indices
to denote sets such as $(\Rv,a,\tau)$, and use bra-ket
notation. For instance,
\BE
\int_0^\b d\tau\sum_\sumind{\Rv,\Rv'}{a,b} V^{\Rv,\Rv'}_{a,b}
\g_{\Rv a}^\star(\tau) \g_{\Rv' b}(\tau)
=\sum_{\mu\nu} V_{\mu\nu} \g_\mu^\star\g_\nu = \bra\g V\ket\g
\EE
Strong-coupling perturbation theory, as formulated in Ref.~\onlinecite{Pairault98,Pairault00}, proceeds first by a so-called Grassmannian Hubbard-Stratonovich transformation, which
amounts to expressing the perturbation $\L\g|V|\g\R$ as the result of a Gaussian integral over auxiliary  Grassmann fields $\psi_{\Rv a}(\tau), \psi_{\Rv a}^\star(\tau)$:
\BE
\label{transfo}
\e^{-\bra\g V\ket\g} = \det V\int [d\psi^\star d\psi] \exp\l[ {\bra\psi V^{-1} \ket\psi + \L\psi|\g\R+ 
\L\g|\psi\R}\r]
\EE
In terms of the auxiliary field, the partition function becomes, up
to a normalization factor,
\BE
\label{z2}
Z=Z_0 \int [d\psi^\star d\psi] \e^{\bra\psi V^{-1}\ket\psi}
\l\L \e^{\L\psi|\g\R+\L\g|\psi\R}\r\R_0
\EE
where $\L\cdots\R_0$ means an unperturbed average over the original electron field $\g$.
Denoting by $\L\cdots\R_{0,c}$ the cumulant averages, and owing to the 
block-diagonality of $H^0$, the above average can be rewritten as\cite{Pairault00}:
\BEA
\label{cum}
\exp
\sum_{R=1}^{\infty}{1\over (R!)^2}\sum_{\Rv\{a_l,a'_l\} }
\int_0^\b\prod_{l=1}^R d\tau_l d\tau'_l &&
\psi_{\Rv a_1}^\star(\tau_1).. \psi_{\Rv a_R}^\star(\tau_R)
\psi_{\Rv a'_R}(\tau'_R)..\psi_{\Rv a'_1}(\tau'_1)\NN\\
&&\times\Big\L\g_{\Rv a_1}(\tau_1).. \g_{\Rv a_R}(\tau_R)
\g_{\Rv a'_R}^\star(\tau'_R)..\g_{\Rv a'_1}^\star(\tau'_1)
\Big\R_{0,{\rm c}}
\EEA
The partition function now takes the familiar form
\BE
\label{z3}
Z\propto \int [d\psi^\star d\psi]  \exp\l\{-S_0[\psi^\star,\psi]-\sum_{R=1}^{\infty} S_{\rm int}^R[\psi^\star,\psi]\r\}\>,
\EE
where the action has a free (Gaussian) part $S_0[\psi^\star,\psi]=-\bra\psi V^{-1}\ket\psi$ and an infinite number of interaction terms
\BE
\label{sint}
S_{\rm int}^R[\psi^\star,\psi]={-1\over (R!)^2}
{\sum_{\{\mu_l,\nu_l\}}}'
\psi_{\mu_1}.. \psi_{\mu_R}
\psi_{\nu_R}^\star..\psi_{\nu_1}^\star
G^{(R)}_{\nu_1..\nu_R\atop \mu_1..\mu_R }.
\EE
\end{widetext}
i.e., the interaction terms are defined in terms of bare vertices given by the exact (connected) electron Green functions $G^{(R)}_{\nu_1..\nu_R\atop \mu_1..\mu_R }$ on a cluster (the primed sum means that all terms share the same cluster index $\Rv$, which is suppressed). These interactions can be treated by the usual diagrammatic perturbation theory, with the unperturbed propagator of auxiliary fermions given by $V$.

In order for the Grassmannian Hubbard-Stratonovitch transformation to be of any use, a connection must be made between the Green functions for the auxiliary fermions and that of the original electrons. This is done in Ref.~\onlinecite{Pairault00}; in particular, the electron one-particle Green function $\Gc$ is related to the self-energy $\Xi$ of the auxiliary fermions by the simple relation $\Gc^{-1} = \Xi^{-1}-V$.
The problem is then reduced to calculating this self-energy $\Xi$, which can be done in perturbation theory using the action (\ref{sint}).

\begin{figure}
\centerline {\includegraphics[width=5cm]{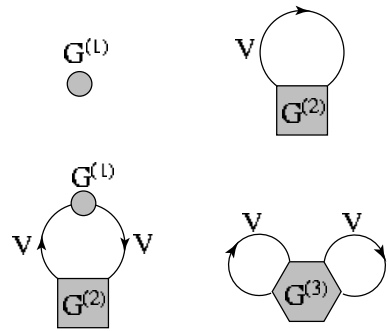}}
\caption{Diagrams associated with the terms of order $V^0$, $V^1$ (above) and $V^2$ (below) in the self-energy $\Xi$ of auxiliary fermions, in strong-coupling perturbation theory. Here only the interaction terms $R=1,2,3$ of Eq.~(\ref{sint}) contribute. The term $R=4$ (i.e., $G^{(4)}$) would come in at order $V^4$.}
\label{diagrams-FIG}
\end{figure}

The simplest term of~(\ref{sint}) corresponds to $R=1$, and is quadratic in $\psi$~:
$S_{\rm int}^1[\psi^\star,\psi] = -\L\psi|G|\psi\R$, where $G_{ab}(\tau)$ is the exactly known one-electron Green function on a cluster. If only this term is kept, then the auxiliary fermion self-energy is precisely $\Xi=G$. If higher interaction terms are considered ($R>1$), then, as illustrated in Fig.~\ref{diagrams-FIG}, propagators of the auxiliary fermions must be inserted and, accordingly, these contributions to $\Xi$ are of higher order in $V$.

Numerically, it is highly impractical to compute anything but the one-electron Green function $G=G^{(1)}$ on a cluster. The latter can be readily obtained by the Lanczos algorithm (see Sect.~\ref{LanczosSEC} below). As explained in Ref.~\onlinecite{Senechal00}, this forces a restriction to the lowest nontrivial order in SCPT. At this order, the one-electron Green function $\Gc$ is given, in operator form, by $\hat\Gc^{-1} = \hat G^{-1}-\hat V$. As a function of frequency, and after restoring orbital and cluster indices, this relation may be expressed as
\BE\label{cpt1}
\hat\Gc(\om) = {\Gh(\om)\over 1-\Vh\Gh(\om)}
\EE
where $\hat\Gc$, $\hat G$ and $\hat V$ stand respectively for the matrices
\BE
\Gc_{ab}^{\Rv,\Rv'}(\om) \quad,\quad
\delta_{\Rv,\Rv'}G_{ab}(\om) \quad\text{and}\quad
V^{\Rv,\Rv'}_{a,b}
\EE

Translation invariance along the superlattice $\Gamma$ allows us to express $V$ and $\Gc$ in terms of a wavevector $\Qv$ belonging to the reduced Brillouin zone (BZ$_\Gamma$, the Brillouin zone of the superlattice) instead of cluster indices. In particular, the hopping term may be expressed as
\BE\label{VQ}
V_{ab}(\Qv) = \sum_\Rv V^{0,\Rv}_{a,b}\e^{i\Qv\cdot\Rv}
\EE
where a superlattice site `0' has been chosen as the origin. In this representation, Eq.~(\ref{cpt1}) becomes
\BE\label{cpt2}
\Gc_{a,b}(\Qv,\om) = \l({\Gh(\om)\over 1-\Vh(\Qv)\Gh(\om)}\r)_{a,b}
\EE
This is the starting formula of Cluster Perturbation Theory on homogeneous systems.

\subsection{Lattices, superlattices and wavevectors}
\label{superlatticeSS}

The Green function $\Gc_{a,b}(\Qv,z)$ is in a mixed representation: direct space within a cluster and reciprocal (Fourier) space between clusters. A pure Fourier representation is preferable, in terms of wavevectors belonging to the Brillouin zone BZ$_\g$ of the original lattice. However, the perturbative treatment breaks translation invariance on $\g$, because it singles out hopping terms between clusters, while translation invariance over the superlattice $\Gamma$ is preserved. As a result, the Green function $\Gc_{a,b}(\Qv,z)$ will depend on two wavevectors $\kv$ and $\kv'$ of the Brillouin zone, except that $\kv-\kv'$ must belong to the reciprocal superlattice $\Gamma^*$. We will demonstrate the following (the frequency argument is suppressed here):
\BE\label{kk'}
\Gc(\kv,\kv') = {1\over L}\sum_{s=1}^L \sum_{a,b=1}^{L} \delta(\kv-\kv'+\qv_s) \Gc_{ab}(\kv)\e^{-i\kv\cdot\rv_a}\e^{i\kv'\cdot\rv_b}
\EE
where (i) $\qv_s$ belongs both to the reciprocal superlattice $\Gamma^*$ and to the original Brillouin zone  BZ$_\g$ ($L$ possible values) and (ii) $\rv_a$ is the position of the lattice site associated with the orbital index $a$.
For simplicity, we will assume a one-band model, so that the orbital index $a$ is also a site index. This assumption will be relaxed in Sect.~\ref{multibandS} below.

The demonstration of Eq.~(\ref{kk'}) is straightforward. We simply apply the basic Fourier transform
\BE
c_{\Rv a} = {1\over\sqrt{NL}}\sum_\kv c(\kv)\e^{i\kv\cdot(\Rv+\rv_a)}
\EE
where $N$ is the number of sites on the superlattice ($N\to\infty$). Applied to the Green function, this yields
\BEA
\lefteqn{\Gc(\kv,\kv') = {1\over NL}\sum_\sumind{\Rv,\Rv'}{a,b} \Gc^{\Rv,\Rv'}_{ab}\e^{-i\kv\cdot(\Rv+\rv_a)}\e^{i\kv'\cdot(\Rv'+\rv_b)} }\NN\\
&=& {1\over N^2L}\kern-0.3em\sum_\sumind{\Rv,\Rv',\Qv}{a,b} \Gc_{ab}(\Qv)\e^{i\Qv\cdot(\Rv-\Rv')}\e^{-i\kv\cdot(\Rv+\rv_a)+i\kv'\cdot(\Rv'+\rv_b)} \NN\\
&=& {1\over L}\sum_{a,b}\sum_\Qv \Gc_{ab}(\Qv)\e^{-i\kv\cdot\rv_a+i\kv'\cdot\rv_b}\delta(\Kv-\Qv)\delta(\Kv'-\Qv) \NN\\
&=& {1\over L}\sum_{a,b}\Gc_{ab}(\Kv)\e^{-i\kv\cdot\rv_a+i\kv'\cdot\rv_b}\delta(\Kv-\Kv') 
\EEA
where the wavevector $\kv$ has been decomposed in a unique fashion as $\kv=\Kv+\tilde\kv$, where $\Kv$ belongs to the reduced Brillouin zone BZ$_\Gamma$ and $\tilde\kv$ belongs to the reciprocal superlattice $\Gamma^*$, and likewise for $\kv'$. Note however that $V(\Kv)=V(\kv)$, since the definition (\ref{VQ}) is invariant with respect to translations of $\Qv$ by a an element of the reciprocal superlattice, such as $\kv-\Kv$; therefore, $\Gc_{ab}(\Kv)=\Gc_{ab}(\kv)$. Since
\BE
\delta(\Kv-\Kv')  = \sum_{s=1}^L \delta(\kv-\kv'+\qv_s)~,
\EE
we recover Formula (\ref{kk'}).

The $\qv_s=0$ term in the above sum is the CPT approximation to the translation-invariant Green function:
\BE\label{cpt3}
\Gc_{\rm CPT} (\kv,z) = {1\over L}\sum_{a,b=1}^{L}\Gc_{ab}(\kv,z)\e^{-i\kv\cdot(\rv_a-\rv_b)}
\EE
This, together with Eq.~(\ref{cpt2}), is the central formula of Cluster Perturbation Theory. In the remainder of this paper, we will drop the index `CPT' on the Green function or related quantities.

In pratice, we are more often interested in the spectral function
\BE
A(\kv,\om) = -2\lim_{\eta\to 0^+}\Imag\Gc(\kv,\om+i\eta+\mu)
\EE
where $\mu$ is the chemical potential. Unless particle-hole symmetry is present, the chemical potential is not known, since the calculation is done at fixed filling. It must be calculated from the density of states (this will be explained in Sect.~\ref{energyS} below). For this reason, it is more pratical to deal with the shifted spectral function
\BE
\label{shiftedA}
\At(\kv,\om) = A(\kv,\om-\mu)=-2\lim_{\eta\to 0^+}\Imag\Gc(\kv,\om+i\eta)
\EE
which does not involve $\mu$ in any way (Note that in the present paper, the Hamiltonian excludes the chemical potential term as well).

The $\qv_s\ne0$ terms are not considered here, but they are a measure of the breaking of translation invariance caused by the different treatment given to intercluster and intracluster hopping. It has been verified that the spectral weight associated with those terms is nonpositive, and that it integrates to zero.

\section{General Remarks}
\label{RemarksS}

\subsection{Limiting cases}
The CPT Green function (\ref{cpt3}) is exact both in the strong and weak-coupling limits. This may look paradoxical at first, since it is a perturbative result that is expected to be exact only when the hopping $V$ goes to zero. However, if we set $U=0$ from the start and use ordinary, `weak-coupling', perturbation theory (i.e. treat inter-cluster hopping as a perturbation of in-cluster hopping), then obviously $V$ is the exact electron self-energy and the basic relation $\hat\Gc^{-1} = \hat G^{-1}-\hat V$ follows.
In other words, and using the notation of a nearest-neighbor Hubbard model, the previous relation may be viewed either as resulting from the lowest order term ($\Xi=G$) in a $t/U$ expansion of the auxiliary fermion self-energy, or from the exact $U=0$ electron self energy ($\Sigma=V$). Needless to say, verifying that the method gives the known $U=0$ spectrum is a practical test that is often carried in practice.
Incidently, Dynamical Mean Field Theory (DMFT) \cite{Georges96} also is exact in those two limits. The relation between the two approaches (DMFT and CPT) remains an open question.

\begin{figure}
\centerline {\includegraphics[width=7cm]{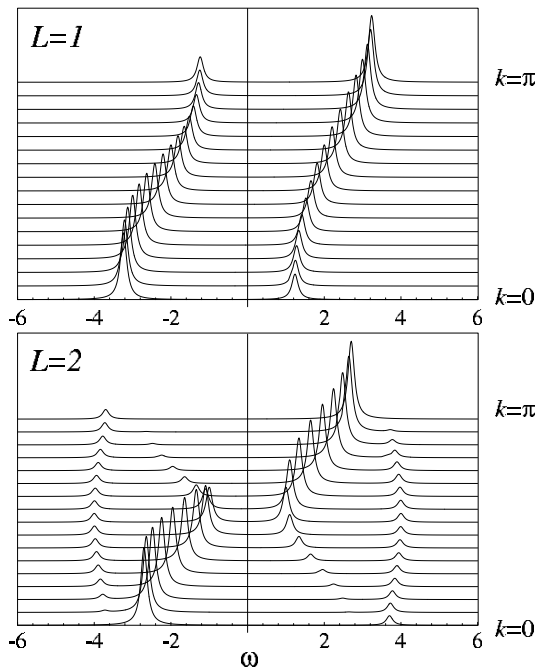}}
\caption{Spectral functions of the one-dimensional Hubbard model ($U=4t$, half-filling) from Cluster Perturbation Theory with $L=1$ (above) and $L=2$ (below).}
\label{1S2S-FIG}
\end{figure}

\begin{figure}
\centerline {\includegraphics[width=\hsize]{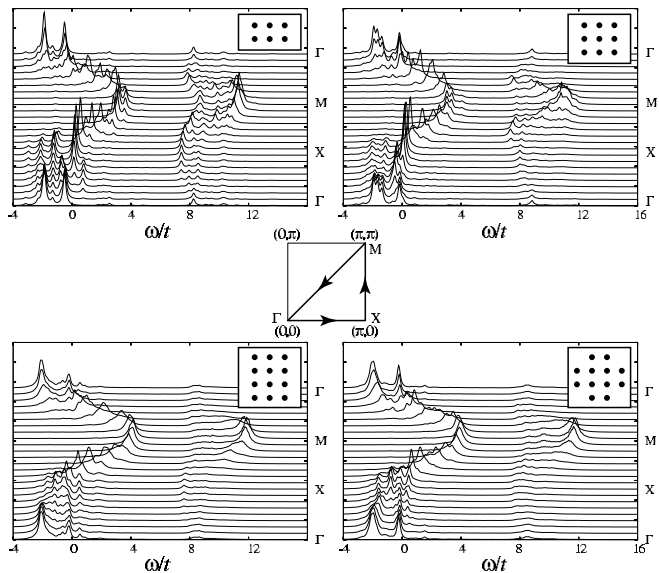}}
\caption{Spectral function of the two-dimensional Hubbard model ($U=8t$, $t'=-0.4t$, $n=2/3$) as calculated by CPT on four different clusters. The momentum scans are indicated.}
\label{diff-amas-FIG}
\end{figure}
\subsection{Nature of the approximation}
It is somewhat difficult to qualify the degree of approximation achieved by Cluster Perturbation Theory. The approach is exact in the limits $U/t =0$, $t/U =0$ and $L\to\infty$. The perturbation parameter is the intercluster hopping $V$ and, according to Sect.~\ref{SCPT-SS} above, the $V$ dependence has been dropped from the auxiliary field self-energy $\Xi$, since only the first diagram of Fig.~\ref{diagrams-FIG} has been kept. However, the CPT Green function itself contains terms of all orders in $V$, as seen from the basic formula (\ref{cpt1}). This is complicated by the presence of the same physical parameter (the hopping $t$) both within the cluster (in which it is treated exactly) and between the cluster (where it is treated perturbatively). In principle, the CPT could be systematically improved by taking into account the higher-order diagrams of Fig.~\ref{diagrams-FIG}, but we have already pointed out that their exact numerical calculation is impractical. In practice, the CPT approximation is controlled most effectively by the cluster size $L$.

A look at Fig.~\ref{1S2S-FIG} is enough to demonstrate the tremendous advantage of treating clusters exactly. Here we treated the one-dimensional, nearest-neighbor Hubbard model at half-filling, with clusters of sizes 1 and  2. If $L=1$, one falls back to the zeroth-order approximation of strong-coupling perturbation theory; this leading order coincides with the Hubbard-I approximation \cite{Hubbard63a}~:
\BE
G(k,\om) = {1\over \om - 2t\cos k- U^2/4\om}
\EE
The difference between this approximation and the atomic limit lies in the dispersion acquired by the two Hubbard bands, and in the momentum-dependent transfer of spectral weight between the two bands. Note that this is also the zeroth-order case of a calculation that was pushed to fifth order (i.e. $(t/U)^5$) in Ref.~\onlinecite{Pairault00}. If we now consider the spectral weight obtained from a cluster of size $L=2$, we already see the main features of the presumed exact result, in particular the weak shadow bands that disperse completely from 0 to $\pi$, i.e., on the reduced (magnetic) Brillouin zone. Of course, the superexchange $J=4t^2/U$ emerges from the exact solution of the two-site cluster and has an impact on the corresponding CPT spectrum.
The $L=2$ spectrum is also closer to the presumed exact result than that obtained by strong-coupling perturbation theory at order $(t/U)^3$ (Fig.~3 of Ref.~\onlinecite{Pairault98}).

In Fig.~\ref{diff-amas-FIG}, we show the spectral function of the two-dimensional Hubbard model (with NN and NNN hopping and density $n=2/3$) for four different clusters (illustrated on the upper right corner of each plot). Minor details may vary from cluster to cluster, but the gross features stay the same~: (i) the general dispersion of the main band; (ii) the two closely separated bands between the  points $\Gamma$ and $X$, the upper one very close to the Fermi level; (iii) the two well-separated ($\Delta\om\sim U$) features at the antifoerromagnetic wavevector $M=(\pi,\pi)$. The larger the cluster, the more poles contribute to the spectral function, making it smoother.

\begin{figure}
\centerline {\includegraphics[width=7cm]{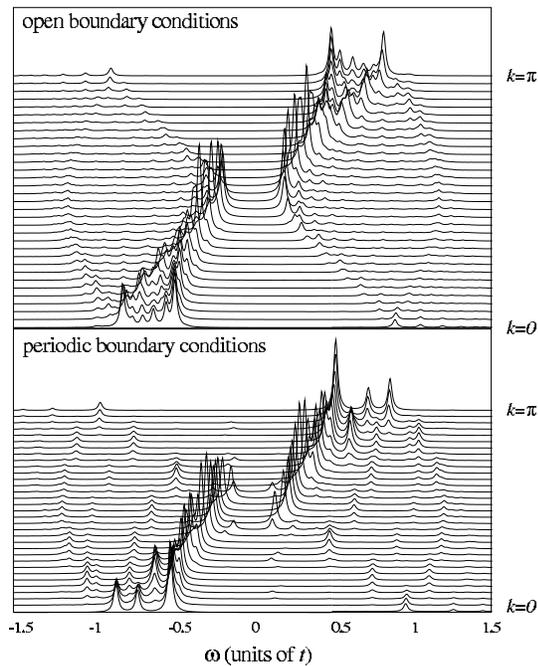}}
\caption{Comparison between the spectral function of the one-dimensional, half-filled Hubbard model with $U=4t$ calculated with open boundary conditions on the cluster (above) and periodic boundary conditions (below), on a 12-site cluster.}
\label{O-P-FIG}
\end{figure}
\subsection{The question of boundary conditions}

In its usual formulation, Cluster Perturbation Theory requires that the cluster Green function be calculated with open boundary conditions, as opposed to periodic boundary conditions. However, this makes the numerics a little more demanding, since the wavevector within a cluster is no longer a conserved quantum number and the Hilbert space to work with is larger by a factor $L$. It has been suggested \cite{Dahnken02} to use periodic boundary conditions anyway, by adding the appropriate hopping terms within the cluster. The same hopping terms are then subtracted within strong coupling perturbation theory, i.e., are added (with the opposite sign) to the perturbation $V$. However, doing this produces spectra that are less accurate than the corresponding spectral obtained from open boundary conditions (Fig.~\ref{O-P-FIG}). Since subtracting the periodic hopping produces a long-range hopping in $V$, this means that CPT apparently does not perform so well with such long-range hopping. Of course, both methods fall back on the exact result in the limiting case $U/t=0$, for the reason described above.

\begin{figure}
\centerline {\includegraphics[width=\hsize]{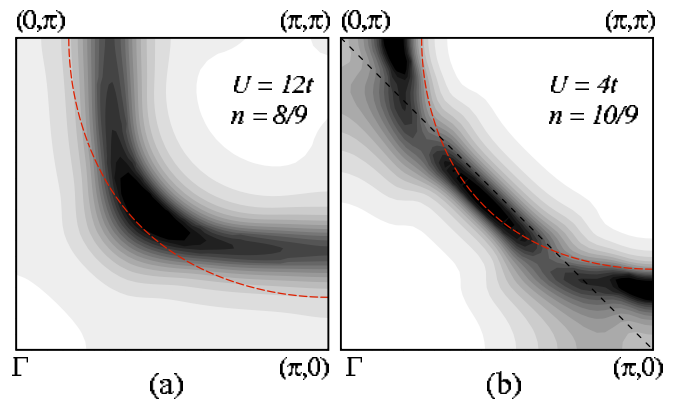}}
\caption{Density plots of the spectral function of the two-dimensional Hubbard model with NNN hopping $t'=-0.4 t$, in the first quadrant of the Brillouin zone. In (a), the density is $n=8/9$ (i.e., 11\% hole doping) and $U=12t$. In (b), $n=10/9$ (i.e., 11\% electron doping) and $U=4t$. In both cases, the noninteracting Fermi surface is shown (dashed curve). The spectral function was integrated over frequency in a interval $\Delta={6\over 25}t$ below the Fermi level.}
\label{pseudo-FIG}
\end{figure}
\subsection{Momentum distribution curves}
One of the key advantages of Cluster Perturbation Theory is the possibility to calculate the spectral function at any wavevector, whereas exact diagonalizations only allow $L$ wavevectors, many of them physically equivalent. This allows the calculation of momentum distribution curves (MDC), i.e., scans of the spectral function at fixed frequency, as can be measured in modern ARPES experiments, or of momentum distribution functions. For instance, Fig.~\ref{pseudo-FIG}a illustrates the spectral function, integrated over frequency in a small interval just below the Fermi level, plotted as a function of wavevector for the two-dimensional Hubbard model with $U=12t$, NNN hopping $t'=-0.4t$ and $n=8/9$. This figure should be compared with ARPES results on Ca$_{2-x}$Na$_x$CuO$_2$Cl$_2$ \cite{Ronning02}, a hole-doped material. One notices a disappearance of the Fermi surface in the antinodal directions $(\pi,0)$ and $(0,\pi)$, which coincides with the opening of a pseudogap in those directions. By contrast, Fig.~\ref{pseudo-FIG}b shows the same spectral function for $n=10/9$ and $U=4t$. There, the Fermi surface disappears first on ``hot spots'' lying on the $(\pi,0)-(0,\pi)$ line; at higher values of $U$, the Fermi surface disappears in the nodal direction, but remains in the antinodal direction, in contrast to the hole-doped case. This should be compared with ARPES results on NCCO\cite{Armitage01}.

\subsection{The $t-J$ model}

The derivation of CPT given above is specific to Hubbard-like models, i.e., models with only on-site repulsion and hopping terms. The $t-J$ model, because it involves correlated hopping, seems excluded from the method. However, CPT has been applied with some success to the $t-J$ model in Refs~\onlinecite{Zacher00, Zacher00B}, and this raises the question as to why it can be successful in this case. No rigorous answer can be given. However, one may first remark that, near the line $J=4t^2/U$, the $t-J$ model is a low-energy limit of the Hubbard model, and therefore CPT wouls be expected to work, at least at low energy. Second, in CPT, the correlated hopping of the $t-J$ model is replaced by ordinary hopping only between clusters, and for a large cluster this may be of little consequence. Finally, it may very well be that the basic formula (\ref{cpt2}) is also the lowest-order approximant to a systematic approximation scheme to the $t-J$ model unknown so far, albeit different from the  Hubbard-model strong-coupling perturbation theory.

\subsection{Finite temperature}

In principle, CPT can be used at finite temperature. Ground state averages are simply replaced by thermal averages (in the canonical or grand canonical ensembles). If the cluster is small enough that all eigenstates can be calculated, then the finite-temperature extension is a straightforward matter. In those cases,  computation time is already short, and a finite temperature is not too much of a numerical burden. For instance, the entropy of the two-dimensional Hubbard model and related thermodynamic quantities can be reasonably well calculated on a $2\times2$ cluster\cite{Roy02}. If the cluster is large enough that the Lanczos algorithm is necessary, then the same techniques used in exact diagonalizations at finite temperature must be used \cite{Jaklic94}. These imply a sampling of initial Lanczos states. In that case computation time may become very important, and some form of parallelization necessary. Work on this topic is in progress.

Incidently, CPT is independent of the way the cluster Green function is calculated. Exact diagonalizations using the Lanczos algorithm are appropriate at zero temperature, but another technique, like Quantum Monte Carlo (QMC), could be used at finite temperature. However, the Green function obtained from QMC is defined in imaginary time, and the continuation to real frequency (via the maximum entropy method) is not only time-consuming, but has problems resolving finer structures in the spectral function. At this point it is not clear whether a Monte-Carlo driven CPT would be a productive method.

\section{The cluster Green function}
\label{clusterS}
%
\subsection{The Lanczos method}
\label{LanczosSEC}

In this subsection, we will briefly review how to calculate the cluster Green function $G_{ab}(\om)$ numerically. First, since the Hamiltonian conserves separately $N_\up$ and $N_\dn$, the numbers of spin-up and spin-down electrons on the cluster, there is a practical advantage in calculating the Green function at fixed values of $N_\up$ and $N_\dn$~: it saves memory (these numbers are not longer conserved once perturbation theory is applied, because of the inter-cluster hopping term). One must first identify the sector that contains the ground state. In the absence of magnetic field, the ground state is expected to be a singlet, i.e., to lie in the $N_\up=N_\dn$ sector. An interesting exception to this rule is indicated by Lieb's theorem\cite{Lieb89}~: on a bipartite lattice with $A$ sites on the first sublattice and $B$ sites on the second sublattice, the ground state of the half-filled, repulsive, nearest-neighbor Hubbard model has spin $|A-B|$. 

The second step is to use the Lanczos algorithm (or the equivalent) to calculate the ground state $|\Om\R$, in the chosen sector. One needs to implement a practical encoding of the states. The Hilbert space can be factorized into spin-up and spin-down parts~: $V = V_\up\otimes V_\dn$. A cluster of $L$ sites with $N_\up=N_\dn=N$ has dimension
\BE
d = \l( {L!\over N!(L-N)!}\r)^2
\EE
For instance, a half-filled 12-site cluster has dimension $d=853~776$, and storing one state in memory (in double precision) requires 6.5~MB. At 16 sites, this figure becomes 1.2 GB, which is impractical unless additional symmetries are used. However, since the method usually demands open boundary conditions, translational invariance is of no use, and  rotational or inversion symmetry are typically too much trouble to implement for what they save in memory. Once a good approximation to $|\Om\R$ has been obtained, one may then proceed to calculate the electron and hole parts of the Green functions~:
\BEA
G_{ab}(z) &=& G_{ab}^e (z) + G_{ab}^h (z) \NN\\
G_{ab}^e (z) &=& \L\Om| c_a{1\over z-H+E_0}c^\dg_b|\Om\R \NN\\
G_{ab}^h (z) &=& \L\Om| c_b^\dg{1\over z+H-E_0}c_a|\Om\R
\EEA
where $E_0$ is the ground state energy and $z$ a complex frequency. In practice, this is done as follows~: one constructs the states $|\phi\R = c^\dg_b|\Om\R$ and $|\phi'\R = c^\dg_a|\Om\R$. Then $|\phi\R$, after being normalized, is used as an initial state in a second Lanczos iteration, which produces a sequence of orthonormal states $|\phi_m\R$, in terms of which the Hamiltonian $H$ is a tridiagonal matrix. At each Lanczos step, the product $\L\phi'|\phi_m\R$ is calculated, and the states $|\phi_m\R$ are not stored, except for the two most recent ones, which are used to calculate the next state. After a sufficient number of iterations (typically a few hundred), one may write with good accuracy
\BE
G_{ab}^e (z) = \sum_m \L\phi'|\phi_m\R x_m \quad,\quad x_m  = \L\phi_m|{1\over z - H+E_0}|\phi\R
\EE
Since $\ket{\phi_0}\propto\ket\phi$, one then has, for a given value of $z$, to solve a tridiagonal system of equations
to find the column-vector $\{ x_m\}$, and this is relatively efficient numerically. The same procedure is repeated for the hole part $G_{ab}^h (z)$. In practice, one stores in memory the tridiagonal matrix $\bra{\phi_m}H\ket{\phi_n}$ for each pair of sites $(a,b)$, as well as the products $\L\phi'|\phi_m\R$. This means that the Lanczos procedure is used only at the beginning, before the sweep over frequencies is performed. 

Note that the full Hamiltonian can easily be stored in memory during the Lanczos procedure. Indeed, it can be expressed as
\BE
H = K_\up\otimes 1 + 1\otimes K_\dn + V_{\rm Coul.}
\EE
where the last term is the (diagonal) Coulomb repulsion, and the first term is the hopping term (or kinetic energy), which acts separately in the space of spin-up and spin-down electrons. The diagonal Coulomb term is stored in memory; this costs the equivalent of one Lanczos vector. The kinetic energy operators $K_\up$ and $K_\dn$ are also stored in memory, but these are sparse matrices of relatively small dimension (typically $\sim\sqrt d$), and their memory cost is negligible.

\subsection{Filling and superclusters}

One of the disadvantages of working with small clusters is the restriction this imposes on the electron density: only a small sample of $L+1$ commensurate fillings is available, in steps $\Delta n=2/L$ (if we set $N_\up=N_\dn$). This may be alleviated to a large degree by the use of superclusters, i.e., larger clusters made of clusters of different fillings. The hopping terms linking the different clusters of the supercluster are treated in perturbation theory to the same degree as above: $(\hat G^{\rm sc})^{-1} = \hat G^{-1}-\hat W$, where $G^{\rm sc}$ is the approximate supercluster Green function, $G$ is the exact cluster Green function (more precisely, the direct sum of the individual Green functions of the clusters making up the supercluster), and $\hat W$ is the inter-cluster hopping matrix. Once $G^{\rm sc}_{ab}$ is known, then the basic CPT formulas (\ref{cpt2},\ref{cpt3}) may be applied to find $\Gc_{\rm CPT}(\kv,z)$. Of course, this procedure does not increase the accuracy of the method, but it allows for a more tunable filling.

\begin{figure}
\centerline {\includegraphics[width=8cm]{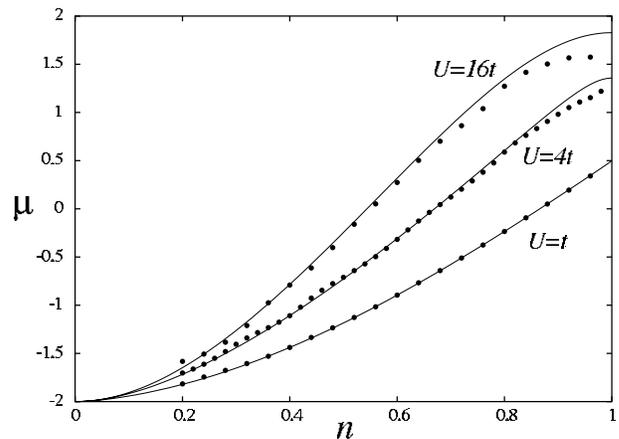}}
\caption{Chemical potential as a function of density in the one-dimensional Hubbard model, as calculated by CPT using a supercluster of ten ten-site clusters (dots). This is to be compared to the exact result obtained by differentiating the exact ground-state energy density with respect to density, following the prescription of Ref.~\protect\onlinecite{Lieb68} (solid line).}
\label{mu-FIG}
\end{figure}

For instance, using ten clusters of ten sites allows us to sweep over electron densities in steps of 0.02. Fig~\ref{mu-FIG} illustrates the chemical potential $\mu$ calculated as a function of density for the one-dimensional Hubbard model. Notice that the accuracy of the method deteriorates at strong coupling as we approach half-filling. This is due to the increasing narrowness of the Hubbard bands as $U/t$ becomes large, which makes the density of states more strongly peaked. This, in turn, makes the calculation of the Fermi level $\mu$ more difficult as the band edge is approached. The numerical evaluation of $\mu$ is discussed further in Sect.~\ref{energyS}.

\section{Multi-band models}
\label{multibandS}

Cluster Perturbation Theory can be applied equally well to multiband Hubbard models, like the three-band of Ref.~\onlinecite{Emery87}, or the four-band model of Ref.~\onlinecite{Andersen94,Andersen95}, meant to describe the Physics of $\rm CuO_2$ planes. Formula (\ref{cpt3}) can be readily adapted to the case of many bands by replacing latin indices $a,b$ by composite indices $(m,\a), (n,\b)$, where $m,n$ are site indices and $\a,\b$ band indices. The latter play a spectator role in the calculations of Sect.~\ref{superlatticeSS}, and the many-band CPT formula reads
\BE\label{cptMB}
\Gc_{\rm CPT}^{\a\b} (\kv,z) = {1\over L}\sum_{a,b=1}^{\norb L}\Gc_{mn}^{\a\b}(\kv,z)\e^{-i\kv\cdot(\rv_m-\rv_n)}
\EE
from which we derive the many-band spectral function
\BE
A^{\a\b}(\kv,\om) =  -2\lim_{\eta\to 0^+}\Imag\Gc^{\a\b}(\kv,\om+i\eta+\mu)
\EE
If an electron is annihilated in a combination of bands by the operator $u_\a c_\a(\kv)$, then the relevant spectral function is (summation over repeated indices is implicit)
\BE
\rho_{\a\b} A^{\a\b}(\kv,\om) \qquad \rho_{\a\b}=u_\a u^*_\b
\EE
More generally, $\rho_{\a\b}$ can be any suitable density matrix (with unit trace) and may correspond to a pure state (like above) or to a mixed state, depending on the physical process under study. Giving it a suitable momentum dependence may also be a way of including matrix-element effects, as in Ref.~\onlinecite{Dahnken01}. But the simplest case would be that of a photoemission process in which the photoelectron comes from any band with equal probability, in which case the density matrix is unity. This scheme was applied in Ref.~\onlinecite{Dahnken02} (albeit using periodic boundary conditions on the cluster) in studying the spectral function of $\rm Sr_2CuO_2Cl_2$.

\begin{figure}
\centerline {\includegraphics[width=\hsize]{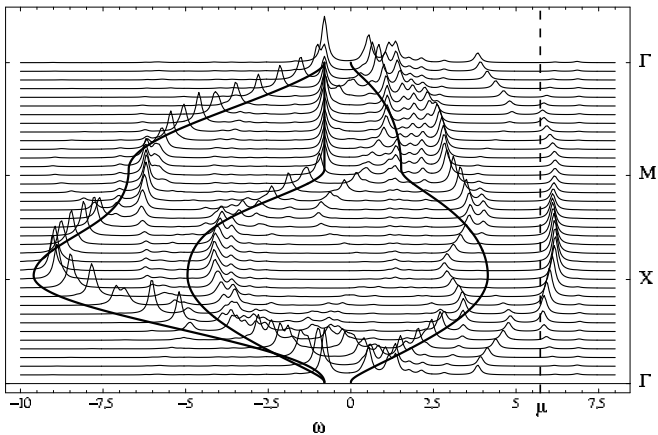}}
\caption{Spectral function for a three-band model of $\rm CuO_2$ planes. The wavevector scan is the same as usual: $\Gamma=(0,0)$, $X=(\pi,0)$ and $M=(\pi,\pi)$. Exceptionally, we plotted the shifted spectral function $\tilde A(\kv,\om) = A(\kv,\om-\mu)$ (the value of $\mu$ is indicated). The $U=0$ dispersion relations are indicated (thick lines).}
\label{cuo2-FIG}
\end{figure}

We illustrate in Fig.~\ref{cuo2-FIG} the spectral function for a three-band Hubbard model. The band parameters are those proposed in Ref.~\onlinecite{Andersen95}: (i) A hopping $t=1.6$~eV between Cu and O orbitals; (ii) a next-nearest-neighbor hopping $t'=1.1$~eV between O orbitals, both diagonally and across a Cu atom; (iii) a  shift $\eps_p=-3$~eV of the O bands with respect to the Cu band. In addition, we put an on-site repulsion $U=4$~eV on the Cu orbital only. This is meant to be a realistic model of  $\rm CuO_2$ planes in YBCO, except for the value of $U$, which we set arbitrarily. Filling was set to 11/12 (i.e. $n=11/6$) for the three bands, which means, in a free electron language, that the upper band has filling 3/4 and the lower two bands are completely filled. We have also plotted the noninteracting ($U=0$) dispersion relations for the three bands (thick lines). In the absence of interaction, the spectrum consists, for each wavevector, of three equal-amplitude delta-function peaks, one for each band. One sees that the interaction has affected mostly the upper band, but that the lower (mostly oxygen) bands are also affected. In particular, the upper band shows extended spectral weight from $\Gamma$ to M, and a gap of order $\sim U/2$ to a narrow, almost dispersionless feature between M and X, lying just above the Fermi level. The lowest band shows a small gap near its inflection point (M), and some portions of the lower bands are somewhat raised by the interaction. This calculation is based on a four-site cluster with three orbitals per site, 25 hopping terms within the cluster and 30 with adjacent clusters.

\section{Energy, double occupancy and chemical potential}
\label{energyS}

In this section we will explain how to calculate the average energy per site and double occupancy from the one-particle Green function, at zero and finite temperature. The basic ingredient of this calculation is the relation between the ground state energy density $\eps$ and the zero-temperature Green function\cite{Fetter71}:
\BE
\eps = -i\int\dk{\kv}\dom \e^{i\om 0^+}(\om+\eps_\kv)
G_\up(\om+i 0^+ \sgn(\om-\mu))
\EE
(we assume here that $G_\up=G_\dn$).
Since we use a complex-frequency Green function throughout, the usual (time-ordered) real-frequency Green function is $G_\up(\om+i 0^+\sgn(\om-\mu))$, whereas the retarded Green function is simply $G_\up^{\rm ret.}(\om) = G_\up(\om+i 0^+)$.
With the help of the spectral representation
\BE
G(\kv,z) = \int\dom {A(\kv,\om)\over z-\om-\mu}
\EE
we find, after integrating around the upper half-plane,
\BE\label{eps}
\eps = \int\dk{\kv}\int_{-\infty}^0\dom (\om+\mu+\eps_k)A(\kv,\om)
\EE
The kinetic energy per site $\Ecin$ may be found in a similar way: starting from 
\BE
\Ecin = 2\int\dk{\kv}  \eps_\kv\L\Om| c_{\kv,\up}^\dg c_{\kv,\up}|\Om\R
\EE
(the factor of 2 comes from the sum over spin), we use the definition of the time-ordered Green function
\BE
G_\s(\kv,t) = -i\L\Om|T c_{\kv,\s}(0)c^\dg_{\kv,\s}(t)|\Om\R
\EE
to express it as
\BEA
\label{Ecin}
\Ecin &=& 2i\int\dk{\kv}\eps_\kv\int\dom \e^{i\om 0^+}G_\up(\om+i 0^+ \sgn(\om-\mu)) \NN\\
&=& 2\int\dk{\kv}\int_{-\infty}^0\dom \eps_k A(\kv,\om)
\EEA
Therefore, by subtracting (\ref{Ecin}) from (\ref{eps}), one finds the potential energy per site in terms of the spectral function:
\BE\label{Epot}
\Epot = \int\dk{\kv}\int_{-\infty}^0\dom (\om+\mu-\eps_k)A(\kv,\om) 
\EE
and the double occupancy is simply $\DO=\Epot/U$.

Before calculating the energy and the double occupancy from formulas (\ref{eps},\ref{Epot}), the chemical potential $\mu$ must be known. But except for special cases with particle-hole symmetry (half-filling, with no NNN hopping), $\mu$ must be calculated from the electron density $n$ by imposing the constraint
\BE\label{mu-n}
n = 2\int\dk{\kv}\int_{-\infty}^\mu\dom \At(\kv,\om) 
\EE
where $\At(\kv,\om)$ is the $\mu$-independent, shifted spectral function of Eq.~(\ref{shiftedA}). In practice, this is done by numerical integration, where the momentum integration is carried before the frequency integration, since the evaluation of the Green function is much faster if momentum sweeps are performed at fixed frequency.

The formulas (\ref{eps},\ref{Epot}) can then be applied to the CPT spectral function $A(\kv,\om)$. One basically needs to compute the two integrals
\BEA\label{I1I2}
I_1 &=& \int\dk{\kv}\int_{-\infty}^0\dom \eps_k A(\kv,\om) \NN\\
I_2 &=& \int\dk{\kv}\int_{-\infty}^0\dom \om A(\kv,\om)
\EEA
and to take thereafter the combinations
\BE
\eps = I_1+I_2+\mu{n\over 2} \qquad \Epot = -I_1+I_2+\mu{n\over 2}
\EE
Some care must be exercised in evaluating these integrals.
The individual peaks of the spectral function being lorentzian curves, they have long tails behaving like $1/\om^2$ at large frequencies. Therefore, the integral $I_2$ above does not exist except in the limit $\eta\to0$. The way out of this problem is to add and subtract a spectral function that integrates analytically and that cancels the leading asymptotic behavior. For instance, on may subtract the atomic limit ($U\to\infty$) result
\BE
{A_0(\kv,\om)\over2\pi} = (2-n)\delta(\om+U/2)+n\delta(\om-U/2)
\EE
from $A(\kv,\om)$ in the integrand (after a lorentzian broadening) and add the corresponding contribution to the result.

\begin{figure}
\centerline {\includegraphics[width=8cm]{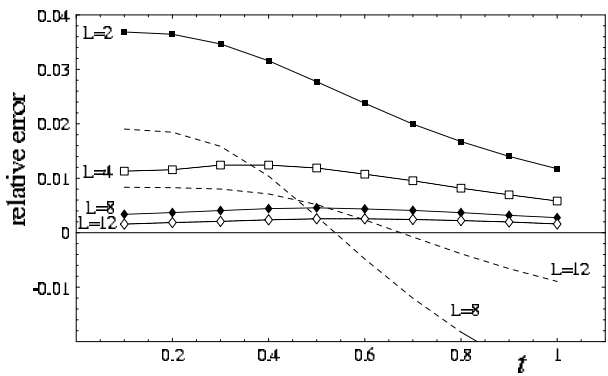}}
\caption{
Comparison (expressed in relative difference) between the ground-state energy density of the half-filled, one-dimensional Hubbard model calculated from the CPT spectral weight (Eq.~(\ref{eps})) and the exact result computed following Ref.~\protect\onlinecite{Lieb68}. The results are displayed as a function of the hopping $t$, for $U=2$ and various cluster sizes $L$ (connected symbols). For comparison, the exact diagonalization values of finite clusters with periodic boundary conditions are also shown (dashed lines).}
\label{ecartE-FIG}
\end{figure}

\begin{figure}
\centerline {\includegraphics[width=8cm]{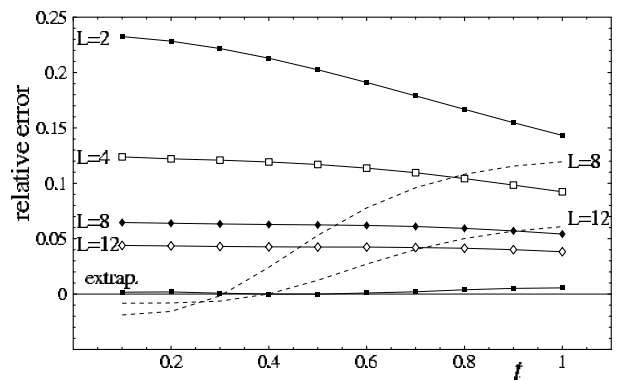}}
\caption{Same as Fig.~\protect\ref{ecartE-FIG}, this time for the double occupancy, calculated from Eq.~(\ref{Epot}). An extrapolation of the results to infinite cluster size ($L\to\infty$) using a quadratic fit in terms of $1/L$ is also shown, and is accurate to within 0.5\%. exact diagonalization results for $L=8$ and $L=12$ are also shown.}
\label{ecartDO-FIG}
\end{figure}

The integrals are then computed numerically from $-\Om_0$ to $0$, where $\Om_0$ is a frequency large compared to the bandwidth or the interaction (typically $\Om_0\sim 10\,{\rm max}(U,t)$). In the improper domain $[-\infty,\Om_0]$, the integral is evaluated using the asymptotic expansion of the Green function
\BE
\Gc(\kv,z) = {1\over z} + {M_1(\kv)\over z^2} +{M_2(\kv)\over z^3} + \cdots 
\EE
where the first moments of the simple Hubbard model are\cite{Kalashnikov72,Nolting72}
\BEA
M_1(\kv) &=& \eps_\kv-\mu+\frac12 Un \NN\\
M_2(\kv) &=& (\eps_\kv-\mu)^2+Un(\eps_\kv-\mu)+\frac12 U^2 n  
\EEA

All these integrals depend on the line broadening $\eta$, and are increasingly difficult to compute as $\eta$ decreases. Extrapolating towards $\eta=0$ from moderate values of $\eta$ (between 0.1 to 0.02) is the method of choice. These tricks are also used in the prior calculation of the chemical potential from Eq.~(\ref{mu-n}).

Fig.~\ref{ecartE-FIG} shows the relative difference between the computed ground-state energy density of the one-dimensional Hubbard model at half-filling and the known exact result computed from Ref.~\onlinecite{Lieb68}. Even with as few as 4 sites per cluster, an accuracy of 1\% is achieved. Clusters of 12 sites bring this down to 0.2\% or less. Notice that the energy density of a bare cluster (no intercluster hopping) with periodic boundary conditions (to eliminate edge effects)
is not nearly as close (dashed lines) to the exact result as the CPT prediction. Fig.~\ref{ecartDO-FIG} shows the same calculation for the double occupancy. Here the accuracy obtained is less impressive (4\% for a 12-site cluster), but a simple quadratic extrapolation as a function of inverse cluster size yields the exact result to within 0.5\%. Moreover, the CPT predictions converge to the exact result in a smoother fashion (as a function of $t$) than exact diagonalization data. Cluster Perturbation Theory is therefore a quite accurate way of computing global quantities like the energy density or the double occupancy.

\section{Conclusion}

Cluster Perturbation Theory is an economical method for calculating the approximate spectral function of Hubbard models. Its main advantage over
exact diagonalizations is the access to a continuum of wavevectors and a better approximation to the thermodynamic limit. This, for instance, allows for an investigation of pseudogap phenomena (Fig.~\ref{pseudo-FIG}).
It is also much faster than Quantum Monte Carlo calculations at very low temperatures and may be formulated directly in terms of real frequencies. Its accuracy has been demonstrated here in the calculation of the ground-state energy of the one-dimensional Hubbard model. In our opinion, it is a method of choice for exploring parameter space and trying to connect real materials to Hamiltonians with local interactions.

\begin{acknowledgments}

We thank F. Ronning for showing us ARPES results prior to publication.
Stimulating dicussions with A.-M.~Tremblay are gratefully acknowledged. This work was supported by NSERC (Canada) and FCAR (Qu\'ebec).

\end{acknowledgments}


\end{document}